\documentclass[twocolumn]{aastex62}

\usepackage{rotating}
\graphicspath{{./}{figures/}}

\accepted{09 Feb 2019}
\submitjournal{ApJ}

\shorttitle{Mass loss in M\,4}
\shortauthors{Tailo et al.}
\begin{document}

\title{Mass loss of different stellar populations in Globular Clusters: the case of M\,4}

\correspondingauthor{Marco Tailo}
\email{marco.tailo@unipd.it; mrctailo@gmail.com}
%APM: metterei solo il mail istituzionale
%MT: no perché lo uso di meno e mi trovo meglio con la mia
%\email{marco.tailo@unipd.it}

\author{M. Tailo}
\affiliation{ Dipartimento di Fisica e Astronomia ``Galileo Galilei'', Univ. di Padova, Vicolo dell'Osservatorio 3, Padova, IT-35122}
\author{A. P.  Milone}
\affiliation{ Dipartimento di Fisica e Astronomia ``Galileo Galilei'', Univ. di Padova, Vicolo dell'Osservatorio 3, Padova, IT-35122}
\author{A. F. Marino}
\affiliation{ Dipartimento di Fisica e Astronomia ``Galileo Galilei'', Univ. di Padova, Vicolo dell'Osservatorio 3, Padova, IT-35122}
\affiliation{Centro di Ateneo di Studi e Attivita’ Spaziali “Giuseppe Colombo” - CISAS, Via Venezia 15, Padova, IT-35131}
\author{F. D'Antona}
\affiliation{INAF- Osservatorio Astronomico di Roma, via di Frascati 33, I-00078 Monteporzio, Italy }
\author{E. Lagioia}
\affiliation{ Dipartimento di Fisica e Astronomia ``Galileo Galilei'', Univ. di Padova, Vicolo dell'Osservatorio 3, Padova, IT-35122}
\author{G. Cordoni}
\affiliation{ Dipartimento di Fisica e Astronomia ``Galileo Galilei'', Univ. di Padova, Vicolo dell'Osservatorio 3, Padova, IT-35122}


\begin{abstract}

In a Globular Cluster (GC), the mass loss during the red-giant branch (RGB) phase and the helium content are fundamental ingredients to constrain the horizontal branch (HB) morphology.
While many papers have been dedicated to the helium %APM2: per recuperare spazio and light elements
 abundance in the different stellar populations, small efforts have been done to disentangle the effects of mass loss and helium content.

We exploit the nearby GC NGC\,6121 (M\,4), which hosts two well-studied main stellar populations, to infer both helium and RGB mass loss. We combine multi-band {\it Hubble Space Telescope} photometry of RGB and main sequence (MS) stars of M\,4 with synthetic spectra to constrain the relative helium content of its stellar populations. We find that the second generation stars in M\,4 is enhanced in helium mass fraction by $\rm \Delta Y = 0.013 \pm 0.002$ with respect to the remaining stars that have pristine helium content. 

We then infer the mass of the HB stars by searching for the best match between the observations and HB populations modelled assuming the helium abundance of each population estimated from the MS.  By comparing the masses of stars along the HB, we constrain the mass loss of first- and second-generation stars in M\,4.
We find that the mass lost by the helium enriched population is $\sim 13$\% larger than the mass lost by the first generation stars ($\rm \Delta \mu = 0.027 \pm 0.006 \ M_\odot$). We discuss the possibility that this mass loss difference depends on helium abundance, the different formation environment of the two generations, or a combination of both.
%\franca{We find that the mass lost by the helium enriched population is $\sim 13$\% larger  than the mass lost by the first generation stars. This huge difference can not be attributed to the tiny difference in the structure of the red giants progenitors, and must be due to different stellar ancillary properties ---e.g. the rotation rate--- of stars born in different formation environments.}
\end{abstract}

\section{Introduction} 
\label{SEC_intro}

50 years and more of study of Globular Clusters (GCs) have not yet reached a full understanding of the variegate description of the morphology of the horizontal branch (HB). It has been easily settled that the `first parameter' governing the HB is the metallicity (iron content [Fe/H] and an associated value of [$\alpha$/Fe]) but the `second parameter' governing the cluster-to-cluster differences at fixed metallicity, remained amply discussed until the end of last century \citep[see e.g.][]{fusipecci_1993}, and includes age \citep{sarajedini_1989}, helium abundance \citep{norris_1981a,norris_1981b}, differences in the red giant branch (RGB) mass loss due to dynamics and/or rotation \citep{fusipecci_1987}.

%The position of a star along the Horizontal Branch (HB) of mono-metallic, single age Globular Clusters (GCs) depends on two parameters, namely helium abundance and mass loss, which affect the mass of a star during the red-giant branch  \citep[RGB, see e.g. the seminal work by][]{dantona_2002}.

In this context, the new century full evidence that nearly all Galactic GCs host multiple stellar populations with probably different helium abundance \citep{dantona_2002} provided a new approach to the problem.
Indeed, helium-rich stars evolve faster than stars with lower helium content ($\rm Y \sim 0.250$). As a consequence, for a fixed age,  they will produce less-massive, hotter HB stars,  which exhibit bluer colours than HB stars with pristine helium abundance \citep[e.g.][]{iben_1984,dantona_2002,dantona_2004}, just as stars experiencing a larger mass loss on the RGB. This helps to solve the mistery of the bimodal HBs, like in NGC\,2808 \citep{catelan_1998}, where indeed populations with different helium have been discovered \citep{dantona_2005, piotto_2007}.

In the case of NGC\,2808, the helium abundance of multiple populations inferred from multiple MSs is a powerful tool to break the degeneracy between helium and mass loss in HB models.
Anyway, adding the contribution of a further parameter in shaping the HB morphology, generally does not resolve the problems. In most papers on HB modelling, age is inferred from the comparison of isochrones with observations of main sequence (MS) and sub-giant branch (SGB) stars, and it is not dependent on the HB morphology. While age can be constrained, both helium content and mass loss must be simultaneously varied to reproduce different HB morphologies. The shortcoming of this approach is a strong degeneracy between these two quantities that cannot be unequivocally constrained especially when the helium variations are small and do not produce significant increase in the HB luminosity level. Small efforts have been done to change this traditional approach.  

An important new tool is now available: recent works, based on multi-band \textit{Hubble Space Telescope} (\textit{HST}) photometry, have constrained the helium content of multiple stellar populations in a large sample of GCs by using MS and RGB stars \cite[][and references therein]{lagioia_2018,milone_2018}, thus providing a solid prospect to break the degeneracy of the parameters involved in the HB morphology.

In this paper we exploit {\it HST}\, multi-band images of the nearby GC NGC6121 (or M4) to infer the age and the average helium abundance of first and second generation (1G, 2G) from MS stars.  These helium determinations will be used to constrain, for the first time, the mass loss of 1G and 2G stars individually, by modelling the HB. 

M4 is one of the most-studied clusters in the context of multiple stellar populations and is an ideal target for our purpose; especially due to the simplicity of its chemical patterns.  High-resolution spectroscopy of red-giant-branch (RGB) stars revealed two distinct groups of stars: a first-generation with lower sodium ([Na/Fe]$\sim$0.1) and high oxygen ([O/Fe]$\sim$0.5) and a second stellar generation enhanced in sodium ([Na/Fe]$\sim$0.45)  with lower oxygen ([O/Fe]$\sim$0.2); see \cite{marino_2008,villanova_2011}. Moreover, the 1G and 2G stars define two distinct sequences that can be followed continuously along the entire colour-magnitude diagram (CMD), from the RGB tip \citep[e.g.][]{marino_2008} towards the bottom of the MS \citep[e.g.][]{milone_2014}. 

The HB of NGC6121 is bimodal and is well populated on both sides of the RR-Lyrae instability strip. Furthermore, high-resolution spectroscopy of HB stars demonstrated that red and blue HB stars belong to the 1G and 2G, respectively, thus providing strong evidence of the connection between the HB morphology and the occurrence of multiple stellar populations \citep{marino_2011,villanova_2012}.

\begin{figure*}
\includegraphics[trim={0 2.8cm 0 1.5cm},width=9cm,height=6cm]{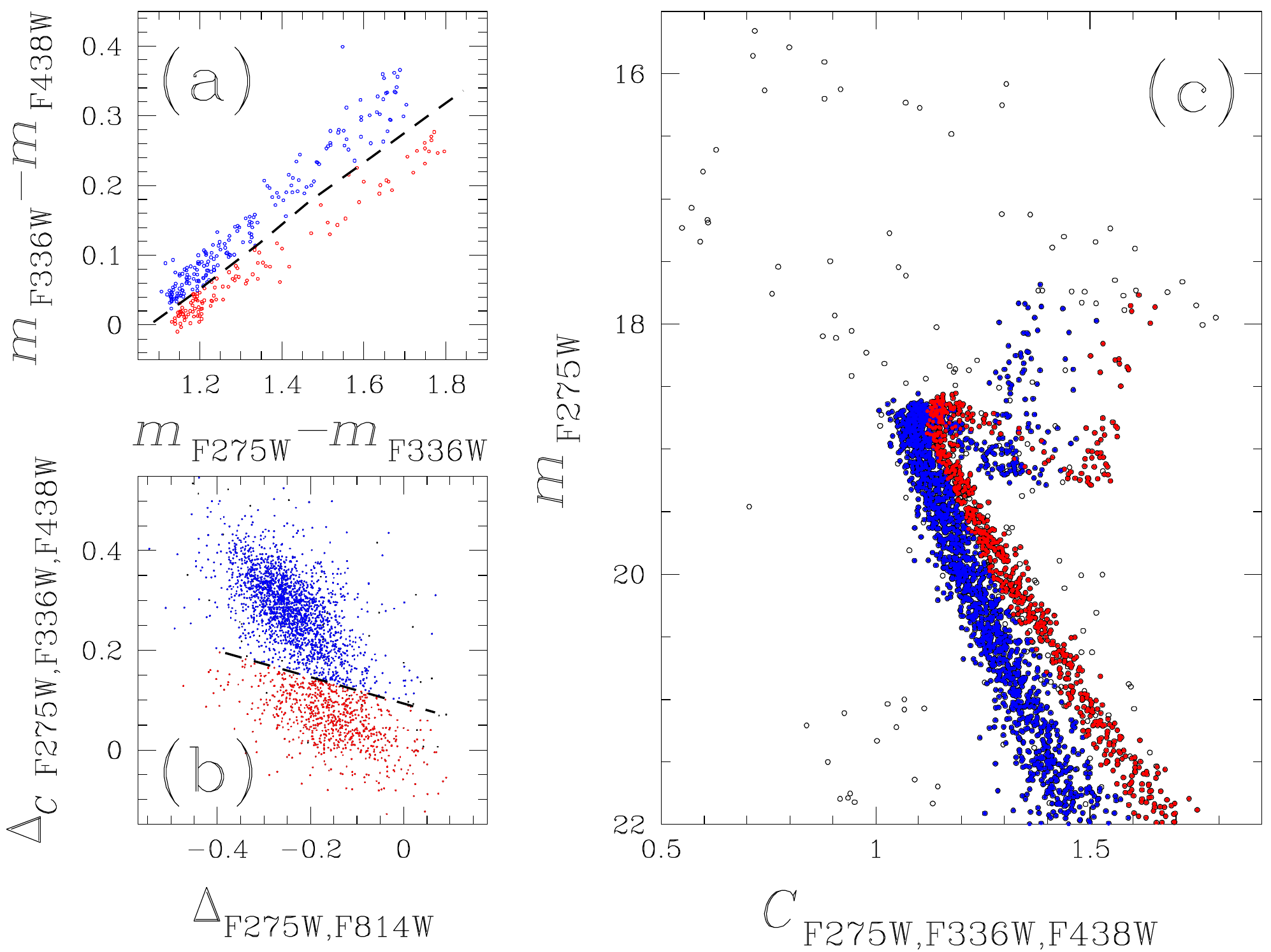}
\vspace{2cm}
\includegraphics[trim={0 2.8cm 0 0cm},width=9cm,height=6.8cm]{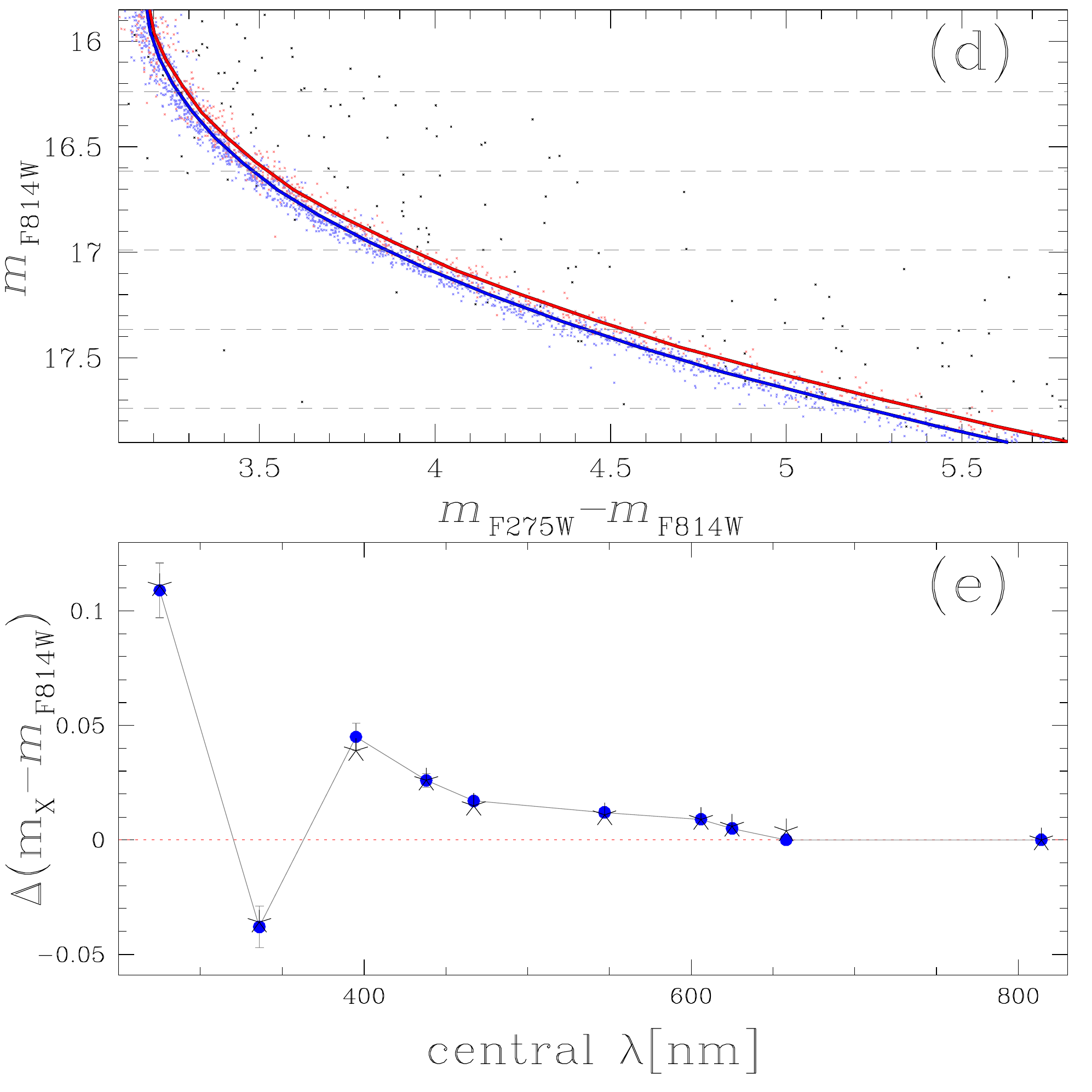}
\caption{This figure illustrates the procedure that we used to identify 1G and 2G stars along the MS and the SGB and to infer their relative helium content. Panels (a) and (b) show the two-color diagram of SGB stars and the ChM of MS stars. The black dashed-dot lines are used to identify the two groups of 1G and 2G stars that we colored red and blue, respectively. Panel (c) shows the $m_{\rm F275W}$ vs.\,$C_{\rm F275W, F336W, F438W}$ pseudo-CMD, while the $m_{\rm F814W}$ vs.\,$m_{\rm F275W}-m_{\rm F814W}$ CMD zoomed on the MS is shown in panel (d). The red and blue lines overimposed on the panel-d CMD are the fiducials of 1G and 2G stars, respectively. The five magnitude values that we used are references to estimate the color seperation between the two MSs and are represented with dashed horizontal lines. Panel (e) shows the $\Delta$($m_{\rm X}-m_{\rm F814W}$) color separation between 2G and 1G stars as a function of the central wavelength of the X filter. 
 }
\label{PIC_MS1}
\end{figure*}

\section{Photometry} 
\label{SEC_dataset}
To identify the 1G and 2G stars along the CMD of M4 and investigate their mass loss, we used literature photometry derived from images collected through the Wide-Field-Channel of the Advanced Camera for Survey (WFC/ACS) and the Ultraviolet and Visual Channel of the Wide Field Camera 3 (UVIS/WFC3) on board {\it HST}. Specifically, we used WFC/ACS photometry in F814W from \citet[see their Table 1]{anderson_2008}; UVIS/WFC3 photometry in F275W, F336W and F438W from \citet[see their Table 1]{piotto_2015}, and photometry collected through the F606W, F625W, and F658N bands of ACS/WFC and the F395N, F467M, F547M bands of UVIS/WFC3 from \citet[see their Table 1]{milone_2018}. Cluster members and field stars have been selected through the analysis of proper motions, as described in more detail in the previous papers; the latter have been excluded from our analysis. 

The photometry in the F275W, F336W, F438W and F814W bands is used to derive the $m_{\rm F336W}-m_{\rm F438W}$ vs.\,$m_{\rm F275W}-m_{\rm F336W}$ two-color diagram of SGB stars and the `chromosome map' (ChM)
\footnote{The ChM is a pseudo two-color diagram that maximizes the separation between 1G and 2G stars. It is constructed by plotting ($m_{\rm F275W}-m_{\rm F336W}$)$-$($m_{\rm F336W}-m_{\rm F438W}$) as a function of $m_{\rm F275W}-m_{\rm F814W}$ for MS and RGB stars. However, it is not a simple two-colors diagram because the sequences are verticalized in both dimensions. We refer to \cite{milone_2015,milone_2017} for a detailed description and to Fig.~1 of \cite{milone_2017} for an illustration of the procedure to derive the ChM.}
of MS stars plotted in panels a and b of Fig.~\ref{PIC_MS1}, respectively,.
These diagrams are used to identify the 1G and 2G stars, which are colored red and blue respectively in Fig.~\ref{PIC_MS1}, while the two groups of 1G and 2G stars along the RGB are identified by using the ChM by \cite{milone_2017}. As an example, Fig.~\ref{PIC_MS1}c shows the selected 1G and 2G stars along the MS, SGB, and RGB of M4 in the $m_{\rm F275W}$ vs.\,$C_{\rm F275W, F336W, F438W}$ pseudo-CMD.
 
\section{The helium abundance of stellar populations in M4} 
\label{SEC_helium}

To infer the relative helium content of 1G and 2G stars, we applied to M4 the procedure introduced by \cite{milone_2012} and used in various papers from our group \citep[e.g.][and references therein]{milone_2018}.
Briefly, we analyzed the CMDs $m_{\rm F814W}$ vs.\,$m_{\rm X}-m_{\rm F814W}$, where X=F275W, F336W, F395N, F438W, F467M, F475W, F606W, F625W and F658N and derived the MS fiducial lines of 1G and 2G stars in each CMD.

We defined five equally-spaced reference points in the magnitude interval $m_{\rm F814W}$ and calculated the corresponding $m_{\rm X}-m_{\rm F814W}$ color differences between the fiducials of 2G and 1G stars, $\Delta(m_{\rm X}-m_{\rm F814W})$.
As an example, in Fig.~\ref{PIC_MS1}d  we over imposed the fiducials of 1G and 2G stars on the $m_{\rm F814W}$ vs.\,$m_{\rm F275W}-m_{\rm F814W}$ CMD  and marked the five reference points with gray dashed horizontal lines.
 We plot in Fig.~\ref{PIC_MS1}e $\Delta(m_{\rm X}-m_{\rm F814W})$ calculated for the available X filters at the reference point $m_{\rm F814W}=17.36$ as a function of the central wavelength of the X filter.

For each reference point, we calculated a reference spectrum and a grid of comparison spectra by using ATLAS 12 and Synthe \citep{castelli_2005a,castelli_2005b,kurucz_2005,sbordone_2007}.
We assumed for the reference spectrum unenriched helium content, Y=0.250, and the individual abundances of C, N, and O inferred for 1G stars by \cite{marino_2008} 
: [C/Fe]=-0.66, [N/Fe]=0.42, [O/Fe]=0.45
; while gravity and effective temperature are taken from the best-fit isochrone. We used for this purpose the isochrones from the database presented in \cite{tailo_2016b} of appropriate metallicity.  We reach a good fit of the CMD with an age of  12.0 Gyr, E(B$-$V)=0.43, $\rm (m-M_{V}) = 11.41$
 and Z=0.002\footnote{These values agree with those provided by \cite{schlafly_2011}, see also: https://irsa.ipac.caltech.edu/applications/DUST/}.
 
Comparison spectra have different abundances of C, N, O and Y. Specifically, [C/Fe], [N/Fe], and [O/Fe] range from $-$1.10 to $-$0.30, from 0.42 to 1.32, and from 0.05 to 0.55, respectively, in steps of 0.1 dex. The helium abundance ranges from Y=0.250 to 0.280 in steps of 0.001.
We used for all the spectra the average abundance of iron, [Fe/H]=$-$1.14, and $\alpha$ elements,  [$\alpha/Fe$]=0.4,  which are consistent with the values estimated by \cite{marino_2008}.

Each synthetic spectrum was convolved with the transmission curves of the filters used in this work to derive the corresponding $\Delta(m_{\rm X}-m_{\rm F814W})^{\rm synth}$ colour difference. We used the Y, [C/Fe], [N/Fe], [O/Fe] values of the comparison spectrum that provides the best fit with the observations, to derive the best-estimate of the relative Y, C, N, and O abundances of 2G and 1G stars.

From the five positions, we find that 2G stars are enhanced in helium by $\Delta$Y=0.013$\pm 0.002$ with respect to the 1G. This result is consistent within one sigma with previous determination based on multi-band {\it HST} photometry of RGB stars, \citep[$\rm \Delta Y = 0.009\pm 0.006 $,][]{milone_2018} and on $U, B, V, I$ ground-based photometry of MS stars \citep[$\rm \Delta Y = 0.020\pm 0.004 $,][]{nardiello_2015}.
%Antonino poi inserisci i numeri giusti tu quando hai il tex in mano
We also find that the 2G stars have $\rm \Delta[C/Fe]=-0.25 \pm 0.15$, $\rm \Delta[N/Fe]= 0.80\pm 0.10$, $\rm \Delta[O/Fe]= -0.35 \pm 0.15$. The resulting value of [(C+N+O)/Fe] does not change within $\rm 0.05\pm0.10\ dex$.

\section{Mass loss and simulated Horizontal Branch} 
\label{SEC_models}

To constrain the mass loss of RGB stars in M4, we compared the observed color and magnitude distributions of HB stars with a grid of synthetic HB stellar population models derived from the tracks calculated by \cite{tailo_2016b}, with constant  [(C+N+O)/Fe] as inferred by \cite{marino_2008}.
The tracks are obtained via the evolutionary code ATON 2.0  \citep{ventura_1998,mazzitelli_1999} and following the recipe by \cite{dantona_2002} and \cite{dantona_2008}.
In a nutshell, each HB track is derived by assuming the same helium core mass of the corresponding star at the RGB tip and decreasing the envelope mass by a quantity equal to the mass loss.

We assumed that the mass loss has a Gaussian distribution with center $\rm \mu_{\rm 1G}$ and dispersion, $\sigma_{\mu}$.
We adopted for the HB stars a mass, $\rm M_{HB} = M_{\rm Tip} - \mu \cdot \sigma_{\mu}$, where $\rm M_{\rm Tip}$ is the stellar mass at the tip of the RGB.
The value of $\rm M_{\rm Tip}$ is provided by the isochrone that we used to fit  the MS stars (see \S~\ref{SEC_helium}), while $\rm \mu$ and $\rm \sigma_{\mu}$ are considered as free parameters. In the following we differentiate these values with the subscript 1G and 2G to indicate the two generations of stars, respectively. The MS and RGB models include a mild mass loss following the \cite{reimers_1975} formulation; the free parameter inside the formula has been set to $\rm \eta_R=0.3$, as described in \cite{tailo_2016b}

\begin{sidewaystable*}
\centering
\caption{The inputs used to obtain the simulations described in the text. Columns list the values metallicity (Z), age, the helium enhancement ( $\rm \Delta Y_{2G,1G}$) between the 2G and 1G stars, the mass at the tip of the RGB for both populations ($\rm M_{Tip;1G}$ and  $\rm M_{Tip;2G}$) and their core mass ($\rm M_{c;1G}$ and  $\rm M_{c;2G}$) , the mean value of mass loss with its spread ($\mu_{\rm 2G},\ \mu_{\rm 2G}, \sigma_\mu$) and the p-value of the KS test for both the colour and magnitude distributions ($\rm KS_1$ and $\rm KS_2$).  }
\begin{tabular}{ccccccccccccc}
\hline
\hline
&Z&Age/Gyr&$\rm \Delta Y_{2G,1G}$&$\rm M_{Tip;1G}/M_\odot$&$\rm M_{Tip;2G}/M_\odot$&$\rm M_{c;1G}/M_\odot$&$\rm M_{c;2G}/M_\odot$& $\mu_{\rm 1G}/M_\odot$& $\mu_{\rm 2G}/M_\odot$& $\sigma_\mu/M_\odot$&$\rm KS_1$&$\rm KS_2$\\
\hline
Sim. 1&$2\times 10^{-3}$ &12.0&0.013& 0.850&0.833&0.482&0.480&0.209 &0.209 &0.006 &$<< 0.01$ &$<< 0.01$ \\
Sim. 2&$2\times 10^{-3}$ &12.0&0.013& 0.850&0.833&0.482&0.480&0.209 &0.236 &0.006& 0.98 & 0.75\\
Sim. 3&$2\times 10^{-3}$ &11.0& 0.013& 0.871&0.852&0.482&0.480&0.228 &0.255 &0.006& 0.98 & 0.87\\
Sim. 4&$2\times 10^{-3}$  &13.0&0.013& 0.832&0.813&0.482&0.480&0.191 &0.217 &0.006& 0.92 & 0.60\\
Sim. 5&$1\times 10^{-3}$ &12.0&0.013& 0.829&0.810&0.490&0.488&0.155 &0.188 &0.006 & 0.90 & 0.65\\
Sim. 6&$3\times 10^{-3}$ &12.0&0.013& 0.868&0.852&0.485&0.483&0.239 &0.264 &0.006 & 0.87 & 0.63\\
Sim. 7&$2\times 10^{-3}$  &12.0&0.040& 0.850&0.806&0.482&0.475&0.209 &0.209 &0.006& 0.94 & 0.80\\
\hline
\hline
\end{tabular}
\label{TAB_inputs}
\end{sidewaystable*}

\begin{figure}
\centering
\includegraphics[trim={0 0 9.5cm 0},width=1\columnwidth]{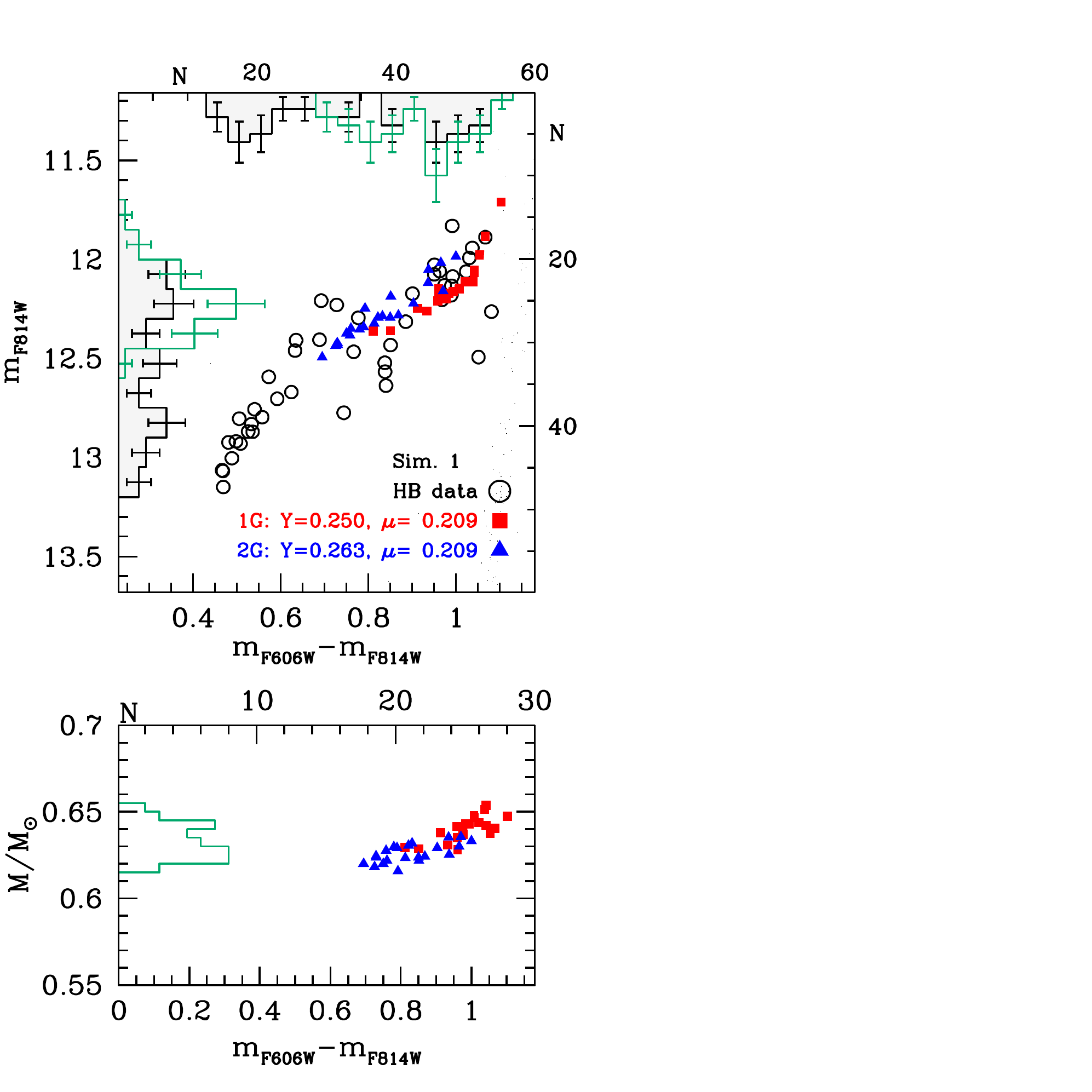}
\caption{\textit{Top:} Comparison of the synthetic HB derived from Sim.\,1 (filled symbols) with the observed HB (open circles). 
Simulated 1G and 2G stars are represented with red squares and blue triangles, respectively.
The color and magnitude distributions of simulated and observed HB stars are represented with aqua and shaded-black histograms, respectively.
\textit{Bottom:} Stellar mass as a function of the $m_{\rm F606W}-m_{\rm F814W}$ color for the simulated stars. The mass distribution is represented by the histogram plotted on the left.}
\label{PIC_results2}
\end{figure}

\begin{figure*}
\centering
\includegraphics[width=1.8\columnwidth]{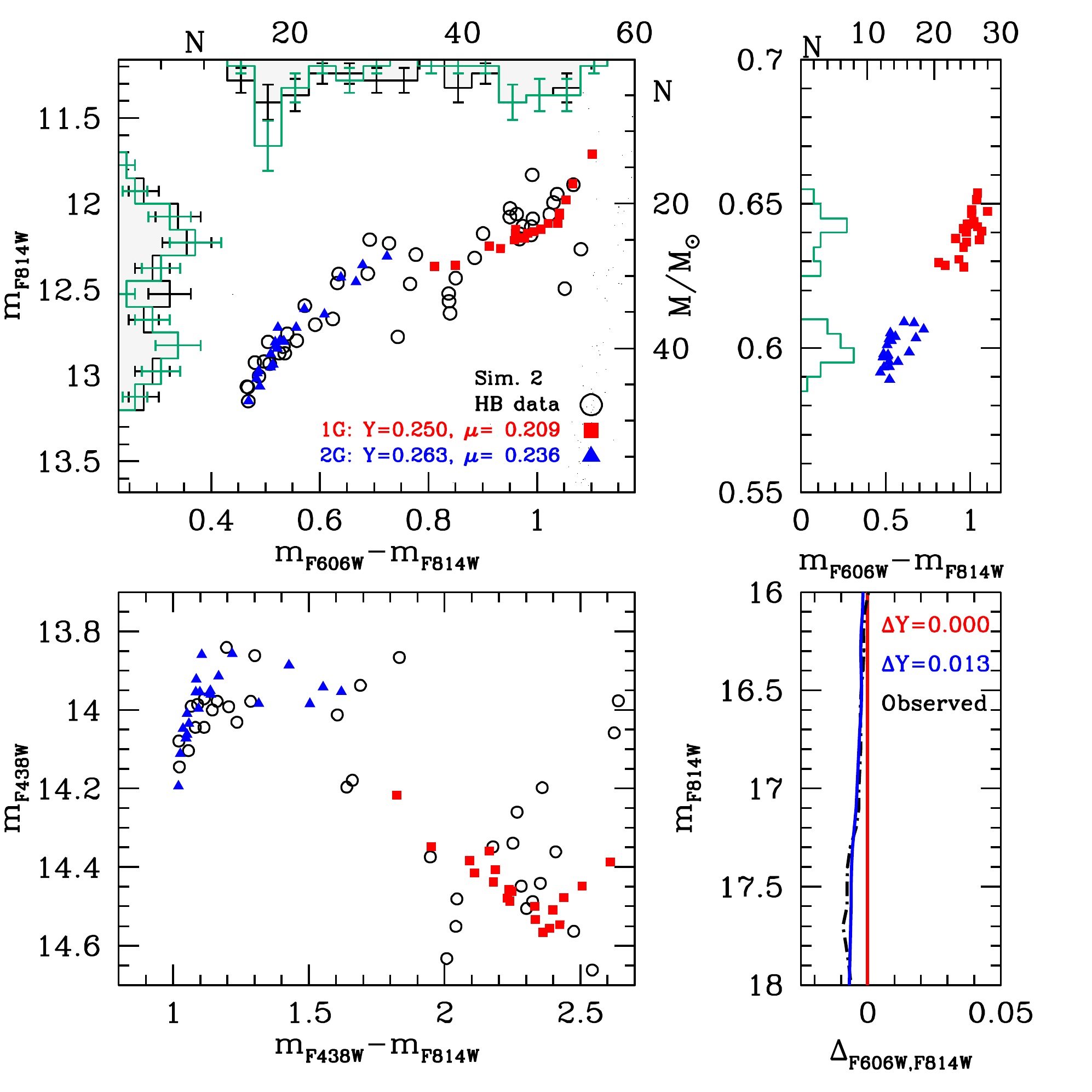}
\caption{\textit{Top.} As in Fig.~\ref{PIC_results2} but for Sim\,2.  \textit{Bottom-left.} Comparison between the observed and the simulated HB in the $m_{\rm F438W}$ vs.\,$m_{\rm F438W}-m_{\rm F814W}$ CMD. \textit{Bottom-right.} $m_{\rm F606W}-m_{\rm F814W}$ color difference between the observed MS of 2G stars and the MS of 1G stars (black dashed-dot line) and color difference between two simulated MSs of 2G and 1G stars that differ in helium content by $\rm \Delta Y=0.013$ (blue line).}
\label{PIC_results3}
\end{figure*}

We assumed that 2G stars are enhanced in helium mass fraction by $\rm \Delta Y_{2G,1G}$=0.013 with respect to the first generation, as inferred from multi-band photometry of MS and RGB stars. Thus the 2G stars have Y=0.263, as we assumed  Y=0.250 for the 1G ones. We generated a grid of models by assuming the same mass loss for both 1G and 2G stars with values ranging from $\rm \mu_{1G}=\mu_{2G}=0.100$  $\rm M_\odot$ to 0.280 $\rm M_\odot$ in steps of 0.001 $\rm M_\odot$. We also included, in each simulation, a spread in mass loss ranging from $\rm \sigma_\mu = 0.000\ M_\odot$ to $\rm 0.020\ M_\odot$ in steps of $\rm 0.001\ M_\odot$. 

The sample of HB stars has been chosen by eye in the F606W--F814W and F438W--F814W CMDs, verifying at the same time that the selected star are consistent with being HB stars in all the analysed CMDs. We identified as red HB stars those redder than $\rm m_{F606W} – m_{F814W} >0.75$, while we identified as blue HB those stars bluer than $\rm m_{F606W} – m_{F814W} <0.75$. We checked that this identification is consistent in the bands we analysed.

 For each individual simulation in the grid, we compared the F606W -- F814W colour distribution of the simulated 1G and 2G HB stars with the observed colour distribution of red and blue HB ones. This choice is justified by the spectroscopic evidence that the red and the blue HB of M4 are mostly populated by stars with lower and higher Na, respectively \citep{marino_2011}.

 The best match between the simulated 1G stars and the observed red HB corresponds to $\rm \mu_{ 1G}=0.209\ M_\odot$ and $\rm \sigma_\mu = 0.006\ M_\odot$, as listed in Tab. \ref{TAB_inputs}.
 In the Sim.\,1 we adopted for 2G stars the mass loss inferred from the first generation.  Figure \ref{PIC_results2} illustrates the results from this simulation, compared to the observed HB (upper panel). Clearly, we obtain a poor fit with the data, as the simulated 2G stars have, on average, redder colours than the observed blue HB. Moreover the p-values obtained from the Kolmogorov - Smirnov (KS) test of the colour and magnitude distribution are both close to zero (see Tab. \ref{TAB_inputs}).  The distribution of mass of the simulated HB, due \textit{only} to the different values of $\rm M_{\rm Tip; 1G (2G)}$ (see Tab.~\ref{TAB_inputs}), is described in bottom panel in Fig.\ref{PIC_results2}. 
 
This attempt shows that it is not possible to reproduce the HB of M4 by assuming the helium difference between 2G and 1G stars inferred from multiple sequences together with \textit{the same mass loss} for both populations. 

To better reproduce the observed colour distribution of both red- and blue-HB stars we used the mass loss of the first generation derived above but assumed that 1G and 2G stars have different mass losses. We generated a grid of models for 2G stars with mass loss values ranging from $\rm \mu_{2G}=0.100\ M_\odot$ to $\rm 0.290\ M_\odot$ in steps of 0.001 $\rm M_\odot$ and with a dispersion ranging from  $\rm \sigma_\mu = 0.000\ M_\odot$ to $\rm 0.020\ M_\odot$ in steps of $\rm 0.001\ M_\odot$. We obtain the best fit between the observed colours of blue HB stars and the simulations of 2G stars for $\rm \mu_{2G}=0.236\ M_\odot$ and $\rm \sigma_\mu = 0.006\ M_\odot$, which is our Sim.\,2 described in Tab. \ref{TAB_inputs}. The comparison of Sim.\,2 with the observations, represented in the left panels of Fig. \ref{PIC_results3}, indicates that a different mass loss for 1G and 2G stars can reproduce the HB of M4, once the different helium abundances for the two populations are constrained from independent features of the CMD.
We thus conclude that to correctly represent the HB stars in M4 the mass loss for the 2G has to be \textit{increased} of $\rm \Delta \mu = 0.027\ M_\odot$. 
In this case the mass distribution of the HB stars exhibits two separated peaks, as reported in the upper right panel of Fig. \ref{PIC_results3}. 

For completeness, in the lower-right panel of Fig.~\ref{PIC_results3} we compare the observed F606W$-$F814W color difference, $\Delta_{\rm F606W,F814W}$ between the MS fiducial lines of 2G and 1G derived in Section~3 (black dashed-dot line) with the corresponding color difference between the 2G and 1G stars from Sim.\,1 (blue continuous line).  This figure confirms that the adopted helium difference between 2G and 1G provides a good match with the observations of MS stars.

\subsection{Impact of observational uncertainties}
Our analysis supports the presence of different mass loss between the 1G and 2G stars on top of their different helium abundance. We test now the impact of the uncertainties of helium abundance, age and metallicity values on this result.

 To investigate the effects of the error in the helium estimate, we repeated the entire procedure with $\rm \Delta Y = 0.015$ and 0.011, the two extreme values obtained in \S~\ref{SEC_helium}. We obtain that with this variations our result changes by $\rm -0.004 M_\odot$ in the case of $\rm \Delta Y = 0.015$ and by $\rm +0.004 M_\odot$ for  $\rm \Delta Y = 0.011$. 

Varying the age by $\rm \pm1~Gyr$ (Sim.\,3 and Sim.\,4 in Tab.\ref{TAB_inputs}) does not vary the $\Delta \mu$ value that best-fits the observed HB. Thus, our result is not significantly affected by age. This is due to the variation of $\rm M_{Tip}$ with age. If we assume a linear relation, we obtain a slope of $\rm \Delta M_{Tip}/\Delta Age=-0.019\ M_\odot/Gyr$ for both the 1G and the 2G models.

When we repeat the procedure using a set of models with a lower metallicity by $\rm \Delta [Fe/H] = -0.15$ ($\rm Z=10^{-3}$) the typical observational uncertainty of spectroscopic analysis (Sim.5, as in Tab.\ref{TAB_inputs}), we obtain $\rm \Delta \mu = 0.033~M_\odot$; a slightly larger value. In the same way with $\rm \Delta [Fe/H] = +0.15$ ($\rm Z=3\times 10^{-3}$, Sim.6, as in Tab.\ref{TAB_inputs}) we obtain $\rm \Delta \mu = 0.025~M_\odot$.
We also obtain $\rm \Delta M_{Tip,1G}/\Delta [Fe/H]=0.123\ M_\odot$ and $\rm \Delta M_{Tip,2G}/\Delta [Fe/H]=0.153\  M_\odot$.

We estimate the error associated to $\rm \Delta \mu$ as the sum in quadrature of the uncertainties introduced by helium, age and metallicity determination. 
We have then $\rm \Delta \mu = 0.027 \pm 0.006\ M_\odot$

\subsection{Is the difference in mass loss necessary? }

As widely discussed, mass loss and helium variations are degenerate parameters in the distribution of stars along the HB. Hence, as an alternative approach to reproduce the observed distributions of HB stars, we used the same mass loss for both the 1G and 2G stars but not the helium difference inferred from the MSs as a constraint. We simulated a grid of synthetic HBs where  $\rm  \Delta Y_{2G,1G}$ ranges from 0.000 to 0.150 in steps of 0.001. The comparison of our grid of HB models with different Y for 2G stars and the observations suggests that the synthetic HB with $\rm \Delta Y_{2G,1G}=0.040$, Sim. 7 in Tab.\ref{TAB_inputs}, provides the best match with data. 

\begin{figure*}
\centering
\includegraphics[width=1.8\columnwidth]{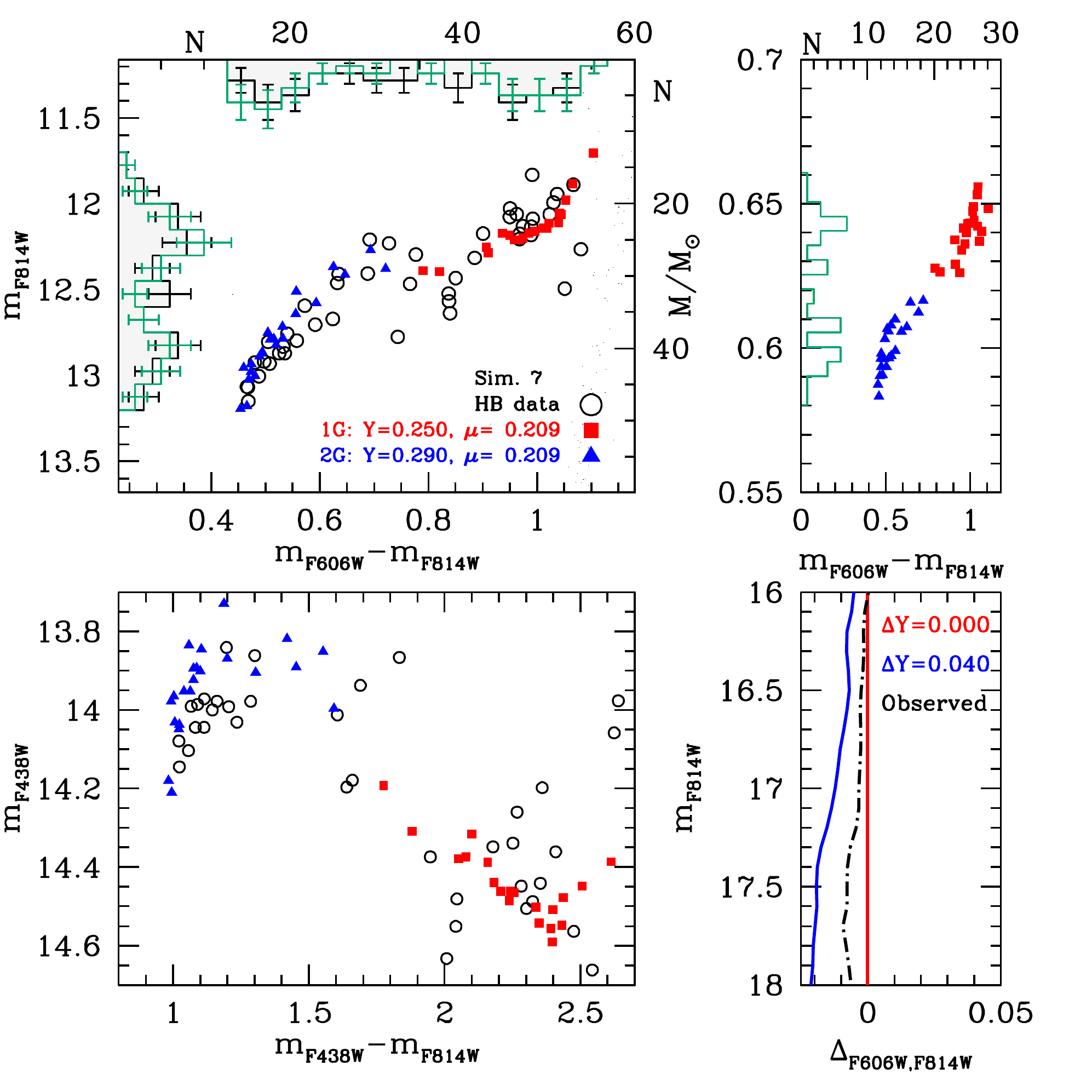}
\caption{As Fig.~\ref{PIC_results3} but for Sim.\,7.}
\label{PIC_results4}
\end{figure*}

The comparison of Sim. 7 with the $\rm (m_{F606W}-m_{F814W})$ vs $\rm m_{F814W}$  and $\rm (m_{F438W}-m_{F814W})$ vs $\rm m_{F438W}$  CMDs is shown in Fig.\ref{PIC_results4}. We note how, differently from Sim. 2 (lower-left panel of Fig.\ref{PIC_results3}), Sim.7 contains a group of blue HB stars more luminous than the observed ones. 

The bottom-right panels of Fig. \ref{PIC_results3} and Fig. \ref{PIC_results4} display the comparison of the fiducials of the observed $\rm \Delta_{F606W,F814W}$ 
(see \S~3) with the theoretical $\rm \Delta_{F606W,F814W}$  obtained from isochrones having the different helium content inferred from Sim.2 and Sim.7, respectively. Clearly, this comparison suggests that an equal mass loss for 1G and 2G stars, as in Sim.7, would result in a too high helium enhancement for 2G stars, not consistent with values inferred from the RGB and the MS.

\section{Summary and discussion}
\label{SEC_conclusion}

The nearby GC M4 is one of the most studied clusters in the context of multiple populations. It hosts two main populations of 1G and 2G stars with different abundances of helium, carbon, nitrogen, oxygen, and sodium that define two distinct MSs, RGBs and asymptotic giant branches \citep[e.g.][]{marino_2008,marino_2017,lee_2010,piotto_2015,lardo_2017}. The HB of M4 is populated on both sides of the RR\,Lyrae instability strip, and the red and blue HB segments are populated by 1G and 2G stars, respectively \citep[e.g.][]{marino_2011,villanova_2012}.  In contrast with more massive GCs that exhibit extended HBs and extreme chemical compositions, M4 is considered a simple cluster in terms of multiple populations. These facts make M4 an ideal candidate to derive the RGB mass loss of different stellar populations in GCs. 

We used multi-band {\it HST} photometry of M4 to infer the relative helium content of its two main stellar populations and to constrain the RGB mass loss.
 The helium abundance was derived by extending the method by \cite{milone_2012,milone_2018} to MS stars of M4. In a nutshell, the ChM was first used to identify 1G and 2G stars along the MS and to derive their colors. Then, we calculated a grid of theoretical stellar atmospheres of MS stars, by assuming different helium and light-element abundance, and finally, we compared the synthetic colors with the observations.
We find that we can match the observations by assuming that 2G stars are enhanced in helium by $\Delta$Y=0.013$\pm 0.002$ with respect to the first generation, which has Y=0.250.

We exploited the helium content of the two populations inferred from multiple MSs to constrain the RGB mass loss.  
To do this we simulate a grid of HBs of 1G and 2G stars with different mass loss and helium and compared the color and magnitude distributions of each simulated HB with the observations. 

By assuming for M4 the values of age, reddening, and distance modulus that provide the best fit between the observed MS, SGB, and RGB and the isochrones, we find that the observations of 1G stars and the red HB are consistent with a mass loss, $\rm \mu_{1G}= 0.209\ M_\odot$.
By using for 2G stars the same helium content as inferred from multi-band photometry of MS stars, the best match between the simulated HB and the observed colors and magnitudes of blue HB stars corresponds to $\rm \mu_{2G}= 0.236\ M_\odot$.
  We conclude that RGB mass loss of 2G stars is larger than that of the 1G ones by $\rm \Delta \mu= 0.027 M_\odot$. 
  
 We have demonstrated that this result is not significantly affected by uncertainties in the adopted metallicity and age.  On the contrary, by assuming the same mass loss value for both 1G and 2G stars, we would need that 2G stars are enhanced in helium by $\rm \Delta Y=0.040$ with respect to the 1G. Such high helium variation is clearly not consistent with the observations of multiple populations along the MS and the RGB.
 
Various studies have shown that it is not possible to reproduce the HB morphology of several GCs, including M3, M13, M14, M22, M92, NGC1851, NGC6388, NGC6441, NGC6363, by assuming the same mass-loss rate for all the stellar populations \citep[see e.g.][]{caloi_2008,joo_2013,tailo_2016b,tailo_2017,vandenberg_2016,denissenkov_2017,vandenberg_2018}.There are also indications that this happens also in Fornax GCs \citep{dantona_2013}.

In this work, we used for the first time the helium abundances of 1G and 2G stars of M4, inferred from multiple MSs, to break the degeneracy between helium and mass loss in HB models and estimate the RGB mass loss. Our conclusion that 2G stars lose more mass than the 1G apparently implies that the RGB mass loss depends on the stellar helium abundance. However, the tiny difference in radius and gravity of the red giant progenitors of the 1G and 2G stars with such a small helium content difference do not physically justify a 13\% difference in the mass loss rate.
As an alternative, we could ascribe this  mass-loss difference to the different formation environments of 1G and 2G stars. 
Indeed, all the proposed scenarios suggest that the 2G forms in the central GC regions \citep[e.g.][]{dercole_2008,valcarce_2011}.
 
The higher density environment may induce a faster initial stellar rotation \citep[see][]{tailo_2015}, which delays the ignition of the helium flash, so that the prolonged evolution at the brightest RGB luminosities can produce a larger total mass loss \citep{mengel_1976,fusi_1978}.
This formation scenario affects the fraction of binary stars that is lower in the 2G as a consequence of the large binary disruption rate in a denser stellar environment \citep{vesperini_2011} and is nicely confirmed by observational work by  \cite{dorazi_2010,lucatello_2015}. 
 
\section*{Acknowledgements }
This work has received funding from the European Research Council (ERC) under the European Union's Horizon 2020 research innovation programme (Grant Agreement ERC-StG 2016, No 716082 'GALFOR', PI: Milone), and the European Union's Horizon 2020 research and innovation programme under the Marie Sklodowska-Curie (Grant Agreement No 797100, beneficiary: Marino). MT and APM acknowledge support from MIUR through the the FARE project R164RM93XW ‘SEMPLICE’ (PI: Milone).
\bibliographystyle{aasjournal}
\bibliography{m4_hbdm} % if your bibtex file is called example.bib

\end{document}